\documentclass[aps,prl,twocolumn]{revtex4}

\usepackage{graphicx}
\usepackage{amsmath}
\usepackage{amssymb}
\usepackage{dcolumn}

\begin{document}

\title{High-Resolution Magnetometry with a Spinor Bose-Einstein Condensate}
\author{M. Vengalattore$^{1}$}
\thanks{These authors contributed equally to this work.}
\author{ J. M. Higbie$^{1*}$}
\author{S. R. Leslie$^1$}
\author{ J. Guzman$^1$} 
\author{L. E. Sadler$^1$}
\author{D. M. Stamper-Kurn$^{1,2}$}
\affiliation{
	$^1$Department of Physics, University of California, Berkeley CA 94720 \\
    $^2$Materials Sciences Division, Lawrence Berkeley National Laboratory, Berkeley, CA 94720}
\date{\today }

\begin{abstract}
We demonstrate a  precision magnetic microscope based on direct imaging of the Larmor precession of a
$^{87}$Rb spinor Bose-Einstein condensate. This magnetometer attains a field sensitivity of 8.3 pT/Hz$^{1/2}$
over a measurement area of 120 $\mu$m$^2$, an improvement over the low-frequency field
sensitivity of modern SQUID magnetometers. The corresponding atom shot-noise limited
sensitivity is estimated to be 0.15 pT/Hz$^{1/2}$ for unity duty cycle measurement. The achieved
phase sensitivity is close to the atom shot-noise limit suggesting
possibilities of spatially resolved spin-squeezed magnetometry. 
This magnetometer marks a significant application of degenerate atomic gases to metrology. 
\end{abstract}

\maketitle

Precision magnetometers that map magnetic fields with high spatial resolution have been
applied to studies of condensed matter systems \cite{CupSC}, biomagnetic imaging \cite{BioMag1}
and tests of fundamental symmetries \cite{EDM2}. Many of these applications require
the measurement of magnetic fields at low ($<$ 10 Hz) frequencies. Current technologies
capable of micron-scale magnetic microscopy include superconducting quantum interference
devices (SQUIDs), scanning Hall probe microscopes, magnetic force microscopes and magneto-optical
imaging techniques \cite{Bending}. Of these, SQUIDs offer the highest sensitivity, demonstrated
at 30 pT/Hz$^{1/2}$ over a measurement area of around 100 $\mu$m$^2$ \cite{SQUID1}.
The low-frequency sensitivity of these devices is limited by $(1/f)$ flicker noise of unknown 
origins \cite{Koch}. 

Magnetic fields may also be sensed by detecting the
Larmor precession of spin polarized atomic gases. To date, atomic
magnetometers have achieved field sensitivities of 0.5 fT/Hz$^{1/2}$
over measurement volumes of 0.3 cm$^3$ \cite{Romalis1}. However,
attaining high spatial resolution with a hot-vapor medium is
precluded by rapid thermal diffusion of the atoms, restricting the
minimum resolved length scale of these magnetometers
to around 1 mm.

Trapped ultracold gases present an attractive medium for a variety
of precision measurements due to their negligible Doppler broadening
and long coherence times \cite{Treutlein, JMH, Jo}.  
Spinor Bose gases, comprised of atoms with non-zero
spin, the orientation of which is free to vary, are particularly well suited to
magnetic microscopy.  
In contrast with hot-vapor atomic magnetometers, the suppression of thermal
diffusion in a gas through Bose-Einstein condensation enables
precise measurements at high spatial resolution.  Also, density-dependent
mean field shifts, which deleteriously affect other types of precision
measurements using dense ultracold gases, do not affect Larmor
precession due to the rotational invariance of interparticle interactions
in a spinor gas \cite{Ho, Ohmi}.

Here, we perform precise magnetic microscopy
with high two-dimensional spatial resolution using a $^{87}$Rb $F=1$ spinor
Bose-Einstein condensate (BEC).  
In our magnetometer, longitudinally spin-polarized spinor
condensates are prepared in an optical trap. Larmor
precession is induced by tipping the magnetization perpendicular to
a bias field imposed along the axis of the condensate. The spins in
each  region of the condensate then precess at a rate that is
proportional to the {\em local} magnetic field. After a variable
integration time, the condensate is probed using
magnetization-sensitive imaging to extract the local Larmor phase.
The {\em difference} in this phase between various regions of the
condensate reveals the spatial variations of the magnetic field.

\begin{figure}
\centering
\includegraphics[width=0.43\textwidth]{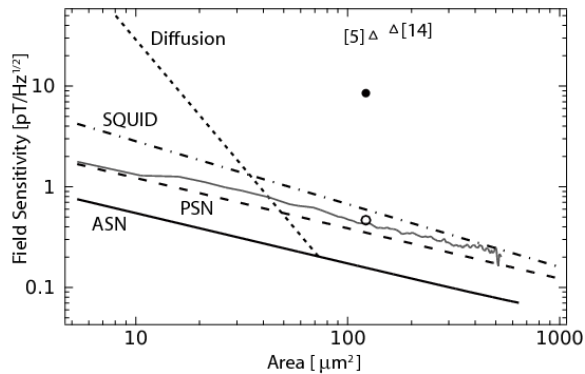}
\caption{Field sensitivity for repeated measurements using the spinor BEC magnetometer.
Curves marked ASN (PSN) represent atom (photon) shot-noise limited sensitivities,
assuming $\tau = 250$ ms, $D = 1$, and the atomic column density and probe
light levels for our experiment (see text). Diffusion of magnetization limits the 
sensitivity for a given length scale by imposing a limit on $\tau$
(short dashed line, assuming $D = 1$). The gray line indicates the measured
spatial root Allan variance; the sensitivity demonstrated in measurements of an applied
localized magnetic field, assuming duty cycles of $D = 0.003$\! ($\bullet$) or $D =1$ \!($\circ$),
is also shown. Results are compared both to the ideal sensitivity of a quantum-limited SQUID
magnetometer (dot-dashed line) and to demonstrated low-frequency sensitivities
 \cite{SQUID1, SQUID6} ($\bigtriangleup$).}
  \label{fig:MagComp}
\end{figure}

The determination of the accrued Larmor phase of a coherent spin
state, such as the transversely magnetized condensate, is subject to
an uncertainty in the initial phase of  $\delta \phi_a = 1/\sqrt{N}$
due to projection noise of measuring $N$ atoms. This noise limits
the field sensitivity over a measurement area $A$ to
$\delta B = \frac{\hbar}{g \mu_B} \frac{1}{\sqrt{\tau D T}} \frac{1}{\sqrt{\tilde{n} A}}$
where $\tau$ is the Zeeman coherence time and $\tilde{n}$ the local
column density of the gas. We assume the measurement is repeated
over a total measurement time $T$ at a duty cycle $D$.  
The $A^{-1/2}$ scaling of field sensitivity with the measurement area
$A$ for the atomic magnetometer may be compared with the area
scaling for SQUID magnetometers. This scaling ranges between $A^{-3/4}$,
for a fixed SQUID sensor coupled optimally to a variable pickup loop, and
$A^{-5/8}$, for direct sensing with a SQUID optimized to operate at
the quantum limit for the noise energy \cite{Koch2}. For either scaling, the
atomic magnetometer outperforms SQUID magnetometers at small
measurement areas (Fig.\ \ref{fig:MagComp}).  


Optical detection of Larmor precession is limited also by photon
shot noise. In this work, the Larmor precession phase is measured by
repeated phase-contrast imaging of the condensate using circular
polarized light \cite{JMH}.  For our probe detuning of $\delta = 2
\pi \times 500$ MHz below the $F=1 \rightarrow F'=2$ (D1) transition
of $^{87}$Rb, the phase contrast signal can be written as $s \simeq
1 + 2 \tilde{n} \sigma_0 (\gamma/2 \delta) [c_0 + c_1 \langle F_y
\rangle]$ where $\sigma_0 = 3 \lambda^2/2 \pi$ is the resonant
cross-section, $\gamma$ is the natural linewidth and $F_y$ is the
projection of the local atomic spin on the imaging axis $\hat{y}$,
which is made perpendicular to the field axis.  The
detuning-dependent constants $c_0 = 0.118$ and $c_1 = 0.274$ describe
the isotropic polarizability and optical activity, respectively.  We
neglect the effects of linear birefringence ($\propto \! \langle
F_y^2\rangle$).  An estimate of the Larmor precession phase is
obtained by tracking the sinusoidal oscillation of the phase
contrast signal across the sequence of phase contrast images. The
photon shot-noise limited sensitivity of this estimate is then
$\delta \phi_\gamma \approx \sqrt{\frac{2}{\eta
N_p}}{\frac{\sqrt{1+\tilde{n} \sigma_0 (\gamma/2 \delta)
c_0}}{\tilde{n} \sigma_0 (\gamma/2 \delta) c_1}}$, giving a field
sensitivity limit of $\delta B = \frac{\hbar}{g \mu_B} \frac{\delta
\phi_\gamma}{\sqrt{\tau D T}}$.  Here, $\eta$ is the
detection quantum efficiency and $N_p$ is the total photon fluence,
integrated across the multi-pulse imaging sequence, within the
region of interest.

For our demonstration,  spin-polarized $^{87}$Rb condensates of up
to $1.4 \times 10^6$ atoms were confined in a single-beam optical
dipole trap characterized by trap frequencies $(\omega_x,
\omega_y, \omega_z) = 2 \pi (165, 440, 4.4)$ s$^{-1}$ \cite{JMH}.
The tight confinement along the imaging axis (condensate radius $r_y
= 2.0\, \mu$m) ensured that the condensate is effectively
two-dimensional with respect to spin dynamics. Larmor precession of
the condensate was induced in the presence a bias field of 165(7) mG
aligned along the long axis of the condensate.  A measurement
integration time of 250 ms was chosen; at longer times,
measurements were hampered by uncontrolled motion of the 
condensate along the weakly confining dimension (see below). 

We operated our ultracold-atom magnetometer under two testing
conditions.  In one, the magnetometer was used to sense the long
length-scale inhomogeneous background magnetic field in our
apparatus.   While these background fields were partially cancelled
by suitable electromagnets, the remaining
background typically varied from shot to shot.  To account for this
fluctuating background, the field profile along the long axis of the
condensate was determined by a third-order polynomial fit to the
magnetometer measurements \cite{backgroundnote}.  The residuals from this fit
were then analyzed to characterize noise limits to our magnetometer.

\begin{figure}
\centering
 \includegraphics[width=0.44\textwidth]{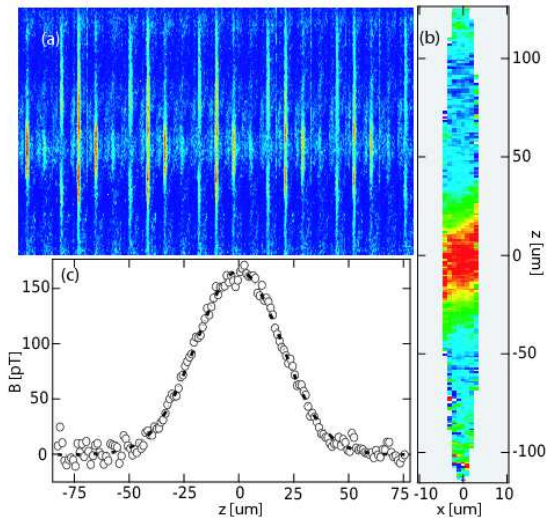}
  \caption{(a) Sequence of phase contrast images of a condensate under the influence of a local optically-induced
Zeeman shift. 
(b) The resulting 2-D map of the magnetic field obtained
by a pixel-by-pixel estimation
of the Larmor precession phase. 
(c) 1-D phase profile showing the
applied field inhomogeneity. The peak strength of the field as
determined by this single-shot measurement is $166.2 \pm 1.2$ pT. }
  \label{fig:LP}
\end{figure}

In the second testing condition, we used the magnetometer to
measure a deliberately applied, localized magnetic
field.  This field was simulated using a circularly
polarized laser beam at a wavelength of 790 nm. The choice of
wavelength and polarization ensured that this beam imposed a local
optically-induced Zeeman shift \cite{Dalibard} on the trapped
atoms (Fig.\ \ref{fig:LP}a). The beam was aligned at an angle
$\theta \sim 60^\circ$ to the direction of the bias field, incident and
focussed in the plane perpendicular to the imaging axis.  The
magnetic background for each run of the magnetometer was again
determined by third-order polynomial fits, but using measurements from
regions far from the localized field. The Zeeman shift due to the localized 
field was extracted from the residual of this fit.

Measurements of this localized field were affected by
small center-of-mass oscillations of the condensate along
its long axis.   An oscillation with amplitude $\delta z$
blurs the magnetic landscape and washes out features comparable to
or smaller than $\delta z$.  Unable to eliminate such
oscillations completely, we monitored the condensate motion for each run of the
magnetometer by a sequence of four images spaced at a quarter period
of the axial trap frequency, taken prior to the Larmor imaging
sequence. We discarded measurements for which an excursion
comparable to the width of the localized field was indicated.

Two-dimensional maps of the magnetic field were obtained from a
pixel-by-pixel analysis of the Larmor precession phase at each
coordinate within the profile of the condensate (Fig.\
\ref{fig:LP}b). The frame-to-frame variation of this signal showed
the characteristic oscillations due to Larmor precession as well as an overall decay
of the condensate number due to off-resonant scattering
of probe light. This decay was taken
into account in obtaining an unbiased estimate of the local Larmor
phase.  Our 2D approach was found to be susceptible to imaging
aberrations, primarily in the narrow ($\hat{x}$) dimension of the
gas.

Alternately, more robust measurements were obtained by integrating
the field measurement over the $\hat{x}$ direction to reduce
measurements to a single resolved direction along the long axis
($\hat{z}$) of the gas.  For this, the aberrated signal profile in
the $\hat{x}$ direction was determined at each $z$ coordinate in the
images from averages over the multiple frames. The phase-contrast
signal height in each image frame and at each $z$ coordinate 
was then determined. 1D  phase profiles were obtained 
as before (Fig.\ \ref{fig:LP}c).

The demonstrated sensitivity of our magnetometer  is shown in 
Fig.\ \ref{fig:MagComp}.  The spatial root
Allan variance from the 1D data was determined for each
of 15 runs of the magnetometer under the first testing
conditions (magnetic background only) and then averaged.  Here, the
measurement \emph{area} is determined by accounting for the
effective $5.3\, \mu$m length over which the aberrated signals are
averaged in the $\hat{x}$ direction.
This noise level matches closely with photon shot-noise
estimates  and is $\sim$3 times that due to atomic shot-noise given
the number of atoms in the corresponding areas. 
Excess noise for areas larger than about 20
$\mu$m$^2$ was found to correlate with the local intensity of the
probe light, an effect we attribute to probe-light induced shifts of
the Larmor precession frequency during imaging. 
This noise can be reduced by using a linearly polarized probe 
with a more homogenous intensity profile and by carefully aligning the magnetic
bias field to be perpendicular to the imaging axis.

\begin{figure}
\centering
\includegraphics[width=0.37\textwidth]{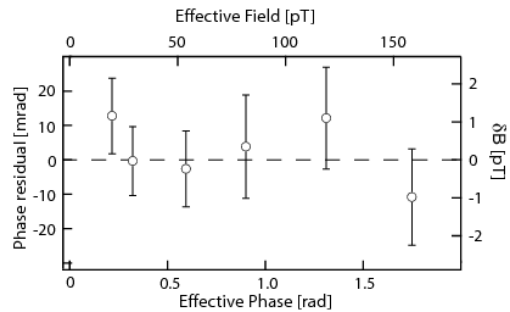}
 \caption{Single-shot measurements of the localized magnetic field
imposed by a laser beam focused to $\sigma=24\, \mu$m rms width. Residuals
from a linear fit, calibrating the field magnitude {\em vs} laser power, are
shown. Error bars indicate standard deviations for 10 measurements per
setting. The measurement time was 250 ms.}
  \label{fig:DR}
\end{figure}

Results from measurements under the second testing condition
(background plus localized field) are shown in Fig.\ \ref{fig:DR}.
Here, the strength of the applied field (peak value of Gaussian
fits) was measured repeatedly at several powers of the
field-inducing laser beam.  From all these measurements, a
calibration between the laser power and the localized field strength
was obtained.  From the residual scatter in the measurements, we
determine the (rms) sensitivity of our Larmor precession phase
measurements as $1.0 \times 10^{-2}$ rad, corresponding to a single-shot field
sensitivity of 0.9 pT over the 120 $\mu\mbox{m}^2$ area under the
Gaussian field profile.  This sensitivity was demonstrated
for field strengths up to 60 pT.  A marginally larger variance for higher
fields points to the existence of small systematic effects,
e.g. \!residual motion of the condensate or variations in the localized 
field strength. Averaging over the entire range of measurements shown
in Fig.\ \ref{fig:DR} yields a phase sensitivity of $1.2 \times 10^{-2}$ rad.

Under repeated operation, our magnetometer, with a low duty cycle of
just $D = 3 \times 10^{-3}$, attains a field sensitivity of 8.3
pT/Hz$^{1/2}$, an improvement over that demonstrated for low-frequency
($<$ 10 Hz) field measurements with modern SQUID magnetometers \cite{SQUID1, SQUID6}.  
Plausible extensions of current cold-atom experimental methods should enable 
duty cycles of order unity.  At full duty cycle, our demonstrated single-shot
sensitivity would correspond to a field sensitivity of 0.5 pT/Hz$^{1/2}$.

As discussed above, in the photon shot-noise limit, the sensitivity of an
atomic magnetometer increases with increasing probe fluence. While
calculations based on linear Raman scattering rates indicated that
reliable phase measurements could be obtained even at a fluence of
3400 photons/$\mu$m$^2$ per frame, it was found that the light induced
losses of the condensate far exceeded those predicted by the calculations.
The discrepancy was attributed to superradiant Raman scattering of
atoms into the $F=2$ hyperfine states, in which atoms would no longer be
observed by our probe. To counter this problem, the superradiant gain was
reduced by lowering the probe intensity. Each frame in the imaging sequence
was obtained by integrating the light from four pulses of light, each of
duration 2.2 $\mu$s and spaced by the Larmor period of $\sim$10 $\mu$s. 
We also increased the motional decoherence of the $F=2$ atoms produced
during superradiance by their preferential scattering of incident light resonant
with the $F=2 \rightarrow F'=3$ (D2) transition. Together, these strategies
enabled a probe fluence of 750 photons/$\mu$m$^2$ per frame. 

Our magnetometry medium, though Bose condensed, is still a gas in
which atoms are free to move. Thus, in determining the phase shift
accrued due to a local magnetic field, one must consider atomic
motion due to both quantum-mechanical and classical
effects. For instance, imposing a weak inhomogeneous
field of characteristic length $\sigma$ leads to quantum diffusive
motion of the fluid.  For times $\tau > \tau_Q = m \sigma^2 /
\hbar$, with $m$ the atomic mass, the motion of the spinor gas will
greatly reduce the phase accrued due to Larmor precession. This
evolution can be considered to be the quantum limit of thermal
diffusion observed in NMR studies \citep{carr54}.  In our
experiment, the condition $\tau = 250 \, \mbox{ms} > \tau_Q$ is
reached at length scales below 10 $\mu$m (Fig.\ \ref{fig:MagComp}).
For integration times $\tau \ll \tau_Q$, effects of quantum diffusion
require that phase measurements $\phi$ be corrected by an amount
$\delta\phi/\phi \sim (\tau/\tau_Q)^4$.

An inhomogeneous magnetic field also exerts forces on magnetic
dipoles. In the extreme case of static inhomogenous fields 
with a Zeeman energy comparable to the chemical potential ($\sim 10^5$ times
larger than those studied in this work), these 
classical forces can modify the density distribution of the condensate \cite{lowresmag}. 
In our case, these forces result in small corrections and limits to the accrued Larmor phase. 
The field strength $B$ and a characteristic length $\sigma$
for its variation define a classical time scale $\tau_C = \sqrt{m
\sigma^2/\mu_B B}$, the time taken by an atom to move $\sigma$ when
accelerated by this field. For an integration time $\tau$, this
classical motion imposes a limit on the maximum detectable phase shift (when
$\tau = \tau_C$) of $\phi_m \simeq \tau_Q/\tau$. 
It should be noted that neither the diffusion of an imprinted phase nor the limitation
on the dynamic range are fundamental; both can be eliminated by
constraining atomic motion, e.g.\ by imposing an optical
lattice potential.

\begin{figure}
\centering
\includegraphics[width=0.46\textwidth]{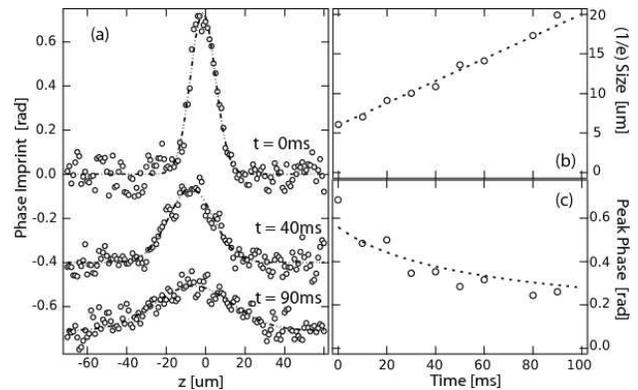}
  \caption{Quantum evolution of an imprinted phase. (a) Cross sections of the phase
imprint after a free evolution time $t = 0, 40$ and $90$ ms. Traces are offset for
clarity. (b) The $1/e$ width of the imprinted phase. The dashed line corresponds to
numerical simulations based on a non-interacting spinor gas. (c) The peak phase
$vs$ the free evolution time. The dashed line corresponds to the expected scaling
imposed by normalization of magnetization.}
  \label{fig:PD}
\end{figure}

To observe the dilution of magnetization due to atomic motion, we imposed a light-induced 
magnetic field localized to a beam waist of 5.4 $\mu$m onto the transversely-magnetized spinor condensate.  
Following a 5 ms exposure to the field-inducing laser beam, the condensate
magnetization was allowed to evolve freely for variable time before
being probed.  During this evolution, the imprinted Larmor phase 
diminished in peak height and grew in extent, matching well with calculations 
based on a non-interacting spinor gas in a localized field (Fig.\ \ref{fig:PD}). 

In conclusion, we have demonstrated a spinor-BEC
magnetometer, a powerful application of ultracold atoms to precision
measurement of scientific and technological significance. Inasmuch
as the Larmor precession phase represents the phase relations among
BECs in several Zeeman states, this magnetometer can be regarded as
a condensate interferometer with high temporal and spatial
resolution. The single-shot phase sensitivity and shot-to-shot
variations of 10 mrad achieved here represent an order of
magnitude improvement over the performance of current BEC
interferometers.

The demonstrated phase sensitivity is close to the atom shot-noise
limit. This augurs spin-squeezed magnetometry via
continuous quantum non-demolition measurements of the condensate
\citep{spinsqueezing} and novel spatially and temporally resolved studies 
of spin squeezed ensembles.
Finally, the high spatial resolution and sensitivity of the spinor
magnetometer render it a powerful detector of quantum spins 
in optical lattices and may facilitate time-resolved, nondestructive
studies of frustration, dipolar interactions and disorder in such systems.

We thank D. Budker and J. Clarke for helpful discussions. This work was
supported by the NSF and the David and Lucile Packard Foundation. 
S. R. L. acknowledges support from the NSERC. 


\begin{references}
\bibitem{backgroundnote} Based on the geometry of the apparatus and the location of magnetic
field sources, we estimate field variations across the condensate due to fourth order
and higher terms to be $\ll$ 1 fT, negligible even at the high precision of the spinor
magnetometer.
\end{references}

\end{document}